\documentclass[twocolumn, showpacs, amsmath, amssymb, prl, superscriptaddress]{revtex4}% Physical Review B
\usepackage{graphicx}% Include figure files
\usepackage{dcolumn}% Align table columns on decimal point
\usepackage{bm}% bold math
\usepackage{color}
\usepackage{epstopdf}
\usepackage[tight]{subfigure}
\usepackage{booktabs}
\usepackage{natbib}

\newcommand{\ph}{\varphi}
\newcommand{\w}{\omega_{\rm p}}
\newcommand{\pha}{\varphi_1}
\newcommand{\phb}{\varphi_2}
\newcommand{\wa}{\omega_{p1}}
\newcommand{\wb}{\omega_{p2}}

\let\oldmarginpar\marginpar
\renewcommand{\marginpar}[1]{\oldmarginpar{\color{red}\footnotesize{#1}}}

\begin{document}

%\newcommand{\note}[1]{\textbf{#1}}

%\title{Magnetic Field Induced Quantum-to-Classical Crossover in a dc-SQUID}
 
\title{Flux Dependent Crossover Between Quantum and Classical Behavior in a dc-SQUID}

\author{S.~Butz}
\email{s.butz@kit.edu}
\affiliation{Physikalisches Institut, Karlsruhe Institute of Technology, 76131 Karlsruhe, Germany}

\author{A.~K.~Feofanov}
\altaffiliation{current affiliation: \'Ecole Polytechnique F\'ed\'erale de Lausanne (EPFL), CH-1015 Lausanne, Switzerland}
\affiliation{Physikalisches Institut, Karlsruhe Institute of Technology, 76131 Karlsruhe, Germany}

\author{K.~G.~Fedorov}
\altaffiliation{current affiliation: Walther-Meissner-Institut, Bayerische Akademie der Wissenschaften, 85748 Garching, Germany}
\affiliation{Physikalisches Institut, Karlsruhe Institute of Technology, 76131 Karlsruhe, Germany} 
\affiliation{National University of Science and Technology MISIS, Moscow 119049, Russia}

\author{H. Rotzinger}
\affiliation{Physikalisches Institut, Karlsruhe Institute of Technology, 76131 Karlsruhe, Germany}

\author{A.~U. Thomann}
\affiliation{Theoretische Physik, ETH Zurich, 8093 Z\"urich, Switzerland}

\author{B.~Mackrodt}
\affiliation{Physikalisch Technische Bundesanstalt, Braunschweig, Germany}

\author{R.~Dolata}
\affiliation{Physikalisch Technische Bundesanstalt, Braunschweig, Germany}

\author{V.~B. Geshkenbein}
\affiliation{Theoretische Physik, ETH Zurich, 8093 Z\"urich, Switzerland}

\author{G. Blatter}
\affiliation{Theoretische Physik, ETH Zurich, 8093 Z\"urich, Switzerland}

\author{A.~V.~Ustinov}
\affiliation{Physikalisches Institut, Karlsruhe Institute of Technology, 76131 Karlsruhe, Germany}
\affiliation{National University of Science and Technology MISIS, Moscow 119049, Russia}
\affiliation{Russian Quantum Center, 100 Novaya St., Skolkovo, Moscow region, 143025, Russia}

\date{\today}

\begin{abstract}

In a coupled system of one classical and one quantum mechanical degree of freedom, the quantum degree of freedom can facilitate the escape of the whole system. Such unusual escape characteristics have been theoretically predicted as ``M\"unchhausen effect''. We implement such a system by shunting one of the two junctions of a dc-SQUID with an additional capacitance. In our experiments, we detect a crossover between quantum and classical escape processes related to the direction of escape. We find that, under varying external magnetic flux, macroscopic quantum tunneling periodically alternates with thermally activated escape, a hallmark of the ``M\"unchhausen effect''.

\end{abstract}

\maketitle

A classical object cannot leave a metastable potential well at zero temperature. In contrast, a quantum object can tunnel out of such a well even at zero temperature. But what happens if a quantum degree of freedom is coupled to a classical degree of freedom? This problem was investigated theoretically in Ref.\,\cite{Thomann09} assuming a weak coupling between the two degrees of freedom. The authors found that the tunneling of the quantum mechanical degree of freedom changes the potential for the classical variable, making it possible for the latter to leave its metastable well. As this effect reminds of the story about Baron M\"unchhausen, who claimed having pulled himself and his horse out of a swamp, the theoretically predicted behavior has been termed ``M\"unchhausen effect''.

The experiments presented in this letter aim at testing the above theoretical idea. We follow the original proposal of using the phase difference across a Josephson junction (JJ) as either quantum mechanical or classical degree of freedom, depending on its effective ``mass" that is proportional to the capacitance of the junction \cite{CL83,LA83}. In our experiment, two JJs with largely different capacitances are coupled by a supercurrent flowing in a superconducting loop that forms a dc-SQUID.

The time evolution of the phase difference $\ph$ across a Josephson junction is analogue to that of a massive particle moving in a washboard potential that is tilted when a bias current is applied. The amplitude of the untilted washboard potential is proportional to the Josephson energy $E_J=\Phi_0 I_c/(2\pi)$, where $I_c$ is the critical current of the JJ and $\Phi_0 = 2.07 \times 10^{-15}$\,Wb the magnetic flux quantum. According to the Josephson relations \cite{JJ65}, the voltage across the JJ is proportional to the velocity $\dot\ph$ of the particle. As long as it is trapped in one of the metastable potential wells, the average voltage is zero. Once the particle escapes its well and runs down the potential, a voltage drop is induced across the junction. This happens at the switching current $I_{\rm sw} < I_c$, when the bias current through the junction, i.e., the tilt of the potential, is strong enough to lower the barrier sufficiently.  
The escape characteristics of JJs have been extensively studied in the past and both the predicted thermal activation (TA) and macroscopic quantum tunneling (MQT) 
processes have been observed \cite{DMC85}.

\begin{figure}[h!]
 \includegraphics[width=.9\columnwidth]{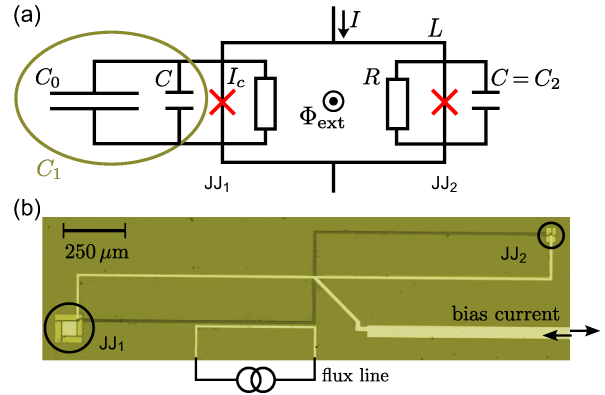}
\caption{(a) Equivalent circuit of the investigated asymmetric dc-SQUID. The parameters of the Josephson junctions are their critical currents $I_c$, normal resistances $R$ and intrinsic capacitances $C$. JJ$_1$ is additionally shunted by a large capacitance $C_0$. $L$ is the loop inductance of the SQUID. (b) Optical microscope picture of the studied dc-SQUID. The bias current lines are fabricated as micro-strips one on top of another in order to minimize their mutual inductances with the gradiometric SQUID loop made as a figure of ``8".}
\label{fig:squid}
\end{figure}

Here we investigate the escape properties of a dc-SQUID consisting of two JJs with deliberately made non-equal capacitances. A sketch of the circuit is shown in Fig.\,\ref{fig:squid}(a). It consists of two nominally identical JJs with critical currents $I_c$ placed in a superconducting loop of inductance $L$. While the dynamics of junction JJ$_2$ is determined by its small intrinsic capacitance $C$, junction JJ$_1$ is additionally shunted with a large on-chip capacitor $C_0 \gg C$. The virtual particle introduced above moves in a two dimensional (2D) potential. The capacitive asymmetry translates into an effectively increased mass of the particle, when it moves in the direction corresponding to the phase difference across junction~JJ$_1$. 

The effect of asymmetry in a dc-SQUID has been of interest for some time. For example, asymmetric damping in a dc-SQUID can be used to improve its noise properties \cite{Rudolph-APL-2012}. Furthermore, SQUIDs with asymmetric critical currents make it possible to suppress phase diffusion \cite{Sullivan-JAP-2013} and are employed for measuring JJ circuits with large quantum fluctuations \cite{Vion-Science-2002,Della-Rocca-PRL-2007}.

\begin{figure}[h!]
 \includegraphics[width=.7\columnwidth]{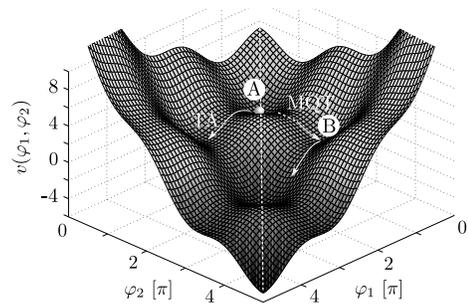}
\caption{2D potential landscape of a dc-SQUID with two identical junctions for $k = 0.15$ at a normalized bias current $j=0.2$ and at $\Phi_{\rm ext} = 0$. The dashed white line indicates the bisectrix where $\pha = \phb$.}
\label{fig:2dpot}
\end{figure}

The equations of motion for the two phase differences $\pha$ and $\phb$ across the junctions are derived using the resistively and capacitively shunted junction (RCSJ) model:
\begin{eqnarray}\label{eq:motion1}
\hspace{-2mm} \frac{1}{\wa^2}\ddot\varphi_1 +\frac{1}{\omega_c}\dot\varphi_1\hspace{-1mm} &=&\hspace{-1mm}- \sin\varphi_1 + j -  k(\pha-\phb- \varphi_{\rm ext})\,,\\
\label{eq:motion2}
\hspace{-2mm}\frac{1}{\wb^2}\ddot\varphi_2+\frac{1}{\omega_c}\dot\varphi_2\hspace{-1mm}&=& \hspace{-1mm} - \sin\phb+j+ k(\pha-\phb - \varphi_{\rm ext})\,.
\end{eqnarray}
Here, $\varphi_i$ is the phase difference across the respective junction JJ$_i$ and $\omega_{pi}=\sqrt{2eI_c/(\hbar C_i)}$ is the corresponding plasma frequency. The damping is defined by $1/\omega_c=\hbar/(2e I_cR)$ and $j=I/(2I_c)$ is the normalized bias current through the dc-SQUID. The coupling of the two junctions is $k = 1/\beta_L = \Phi_0/(2\pi L I_c)$, and $\varphi_{\rm ext} = 2\pi\Phi_{\rm ext}/\Phi_0$ is the phase associated with an externally applied flux $\Phi_{\rm ext}$. Integrating the right-hand side of Eqs.\,(\ref{eq:motion1}) and (\ref{eq:motion2}) with respect to both phase differences yields the equation for the normalized 2D potential:
\begin{eqnarray}
\label{eq:pot}
 v(\pha,\phb)&= &2-\cos \pha - \cos\phb \nonumber\\ 							%%= \frac{U(\pha,\phb)}{E_J} \nonumber \\
&-& j(\pha+\phb) + \frac{k}{2}(\pha-\phb- \varphi_{\rm ext})^2.
\end{eqnarray}
The potential is illustrated in Fig.\,\ref{fig:2dpot}. Due to the weak coupling, there are several minima. A classical particle that escapes out of one of those minima follows a trajectory that leads over the lowest barrier \cite{lefevredevoret92}. Similarly, a quantum particle leaves the minimum via tunneling under the lowest barrier \cite{han02}. Classical (TA) and quantum mechanical (MQT) escape trajectories are illustrated by white arrows in Fig.\,\ref{fig:2dpot}.

An optical microscope picture of the dc-SQUID is shown in Fig.\,\ref{fig:squid}(b). The circuit was fabricated using a submicron Nb/AlO$_x$/Nb trilayer process \cite{dolata05}. A gradiometric design was used in order to suppress unwanted background noise. The SQUID has a loop inductance $L = 4$\,nH, which is distributed symmetrically between the two arms. The two junctions are located at the corners and marked by circles in Fig.\,\ref{fig:squid}(b). Their parameters are given in Tab.\,\ref{tab:para}. The critical current $I_{c, \mathrm{nom}}$ stands for the nominally designed critical current and $I_{ci}$ represents the actual critical current extracted from measurement. Details on how the actual (non identical) critical currents were determined are given later. The capacitances, plasma frequencies and the crossover temperatures are calculated from the measured parameters. The crossover temperature $T_{\mathrm{cr}} = \hbar \omega_p/(2\pi k_{\rm B})$ \cite{tinkham} as given in Tab.\,\ref{tab:para} is the temperature at which quantum and thermal escape rates are approximately equal at zero bias current.

\begin{table}[b]
\begin{tabular}{p{0.21\columnwidth}p{0.12\columnwidth}p{0.12\columnwidth}p{0.10\columnwidth}p{0.10\columnwidth}p{0.14\columnwidth}p{0.14\columnwidth}}
\hline
\hline
 & $I_{c, \mathrm{nom}}$ \newline [nA] & $I_{c}$ \newline [nA] & $R$ \newline[k$\Omega$]& $C$ \newline[fF] & $\w/(2\pi)$ [GHz] & $T_{\mathrm{cr}}$ \newline[mK] \\
\hline
JJ$_1$ (TA) &  560	&465&3.6&1000&6&46\\
JJ$_2$ (MQT)      &560 &515    & 3.6& 3   &115&880\\
\hline\hline
\end{tabular}
\caption{\label{tab:para} Parameters of the two junctions of the dc-SQUID. $I_{c, \mathrm{nom}}$ refers to the nominal value of the junction, while $I_{c}$ is the fluctuation-free critical current extracted from measurements. The capacitances of the junctions as well as their respective plasma frequencies and the cross-over temperatures at zero bias are given by $C$, $\w/(2\pi)$ and $T_{\mathrm{cr}}$, respectively.}
\end{table}
The sample was mounted in a dilution cryostat with a base temperature of $T \approx 25$\,mK. Extensive filtering of the biasing lines was used in order to reduce high frequency noise. Current ramp measurements \cite{wallraff03} were performed for different temperatures to obtain the flux dependent switching characteristics of the asymmetric dc-SQUID. The magnetic flux $\Phi_{\rm ext}$ was applied by sending current through the on-chip flux line as shown in Fig.\,\ref{fig:squid}(b). 

\begin{figure}[tb]
	\includegraphics[width = 1\columnwidth]{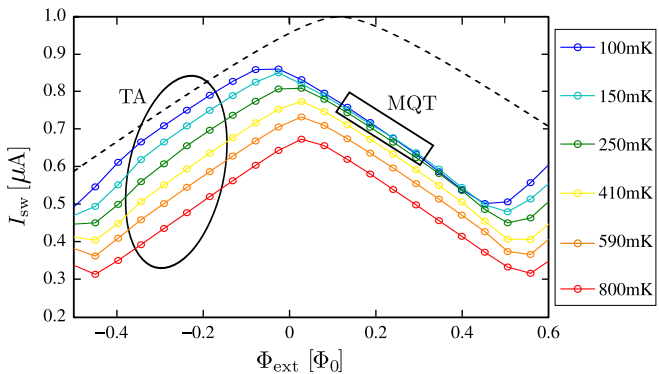}
\caption{Temperature and flux dependence of the mean switching current. Circles indicate a measured value, the lines are guides to the eye. The black oval and rectangle indicate data points used for the fitting of TA and MQT rates, respectively. The dashed line is the fluctuation free critical current of the dc-SQUID calculated from the measured parameters in Tab.\,\ref{tab:para}.}
\label{fig:flux_temp}
\end{figure}

The flux dependence of the mean switching current $I_{\mathrm{sw}}$ is displayed in Fig.\,\ref{fig:flux_temp} for different temperatures ranging from 100\,mK (blue) to 850\,mK (red). Additionally, the fluctuation free critical current calculated from the measured SQUID parameters is shown by a dashed line.  Its maximum is shifted from $\Phi_{\rm ext}=0$ because of the asymmetry in the measured critical currents, see Tab.\,1.
On the left hand side (positive slope) the lowest temperature curve (blue) is close to the zero temperature prediction (black dashed line) and a steady decrease of the mean switching current with increasing temperature is observed, consistent with TA predictions. In contrast, on the negative slope the mean switching current is considerably suppressed compared to the dashed line and independent of temperature between $100$\,mK and $270$\,mK. This is a clear indication that the escape on the negative slopes is induced by quantum rather than thermal fluctuations.

This asymmetric dependence of switching current on flux is explained by the direction dependent escape characteristics shown in Fig.\,\ref{fig:2dpot}. At zero bias current and zero flux, the virtual particle describing the state of the system is trapped in the main minimum (location A) which is symmetric with respect to the bisectrix (white dashed line). At temperatures close to zero, the escape along the classical $\pha$-direction is allowed only under the condition that the barrier is low, i.e $j\approx 1$. However, tunneling in $\phb$-direction is possible while the barrier height in that direction is still appreciable. Therefore, as the bias current is increased from zero, the particle leaves well A in the quantum mechanical $\phb$-direction and escapes passing through the side well indicated by~B.

When increasing the external flux from $\Phi_{\rm ext}=0$, a circulating current is induced in the SQUID loop. It is oriented such that the net current through JJ$_1$ is decreased and through JJ$_2$ increased. Thus, the barrier along $\phb$ is decreased and escape due to MQT is possible already at smaller bias currents. However, at the same time the barrier of the side well B in $\pha$-direction increases until it blocks the motion of the particle. From this point on ($\Phi_{\rm ext}\lesssim \Phi_0/2$), the bias current has to be increased again until the particle can leave the side well B by TA and start running down the potential.  

Once the external flux is turned negative from zero ($\Phi_{\rm ext}<0$), the direction of the circulating current is reversed. Therefore, the barrier of the main well A in the quantum mechanical $\phb$-direction grows while the barrier in the classical $\pha$-direction decreases. The switching current due to MQT increases until it reaches its maximum (still close to zero flux) and  classical TA becomes dominant. From here, the switching current of the dc-SQUID decreases again and follows the well known flux dependence (cf. Ref.\,\cite{lefevredevoret92}). The barrier in the classical direction decreases up to the flux $\Phi_{\rm ext} \approx -\Phi_0/2$ and so does the switching current of the SQUID. In the following, we corroborate this picture by demonstrating the consistency of the data with predicted TA and MQT decay rates, respectively.

The mean switching current shown in Fig.\,\ref{fig:flux_temp} is extracted from switching current histograms which in turn are related to escape rates \cite{fulton74}. In a weakly coupled dc-SQUID the non-trivial potential landscape has to be taken into account in the determination of both classical (TA) and quantum (MQT) decay rates. An expression that takes the two-dimensionality of the thermal decay process into account has been put forward in Ref.\,\cite{lefevredevoret92},
\begin{equation}
\label{eq:gamma_TA}
	\Gamma_{\mathrm{TA}} = \frac{\w}{2\pi}\frac{\omega_{w\perp }}{\omega_{s\perp}}\exp\left(-\frac{\Delta U}{k_B T}\right).
\end{equation}
where the current dependent potential height $\Delta U$ and plasma frequency $\omega_p$ additionally depend on the direction $\theta$ of the decay \cite{lefevredevoret92}. The  angle $\theta$ is measured with respect to the $\varphi_1$-axis and changes with applied flux $\Phi_\mathrm{ext}$. In addition, here we have to account for the asymmetric capacitance by introducing the effective capacitance $C_\mathrm{eff} =C_1C_2/(C_1 \sin^2\theta + C_2\cos^2\theta)$ into the expression for $\omega_p$. Assuming weak coupling, $C_{\rm eff}$ can be calculated from the second derivative of the potential along
the direction of escape. For our weakly coupled SQUID, the ratio of transverse frequencies $\omega_{w\perp}/\omega_{s\perp} $ in the well and around the saddle can be set to unity.

\begin{figure}[b]
 \includegraphics[width = 1\columnwidth]{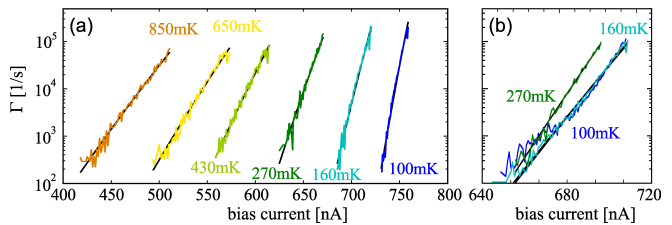}
 \caption{(a) Thermal escape rates at $\Phi_{\rm ext}/\Phi_0 = -0.24$. Measured rates are in color, fit results for the respective temperatures in black. (b) Comparison of measured (color) and calculated (black) MQT rates at $\Phi_{\rm ext}/\Phi_0 = 0.24$. The experimental data is shown for the three lowest temperatures.}
 \label{fig:fitratesTA}
\end{figure}

Equation\,(\ref{eq:gamma_TA}) is used to fit the measured escape rates corresponding to the data points in the black oval in Fig.\,\ref{fig:flux_temp}. It yields a mean critical current $I_0= 490 \pm 9$\,nA and  a current asymmetry paramter $\alpha =  0.053 \pm 0.0004$, such that $I_{c1} = (1-\alpha)I_0 = 465$\,nA and $I_{c2} = (1+\alpha)I_0 = 515$\,nA. This result is in agreement with the designed value of $I_{\rm c, nom} = 560$\,nA within the unavoidable uncertainties in the fabrication process.

The effective junction temperature $T$ is used as an additional fit parameter. The mean value of the four values of $T$ resulting from measurements at different flux values $\Phi_{\rm ext}$ is determined and the standard deviation found to be less than 6\%. The resulting mean temperatures (as given in Figs.\,\ref{fig:flux_temp} and \ref{fig:fitratesTA}) reflect the electron temperatures of the junctions and are, as expected, higher than the measured cyrostat temperature. In Fig.\,\ref{fig:fitratesTA}(a), the measured escape rates (in color) and the fit results (black) are presented for a flux value of $\Phi_{\rm ext}/\Phi_0 = -0.24$.

The decay via quantum tunneling has to account for the two-dimensionality of the situation as well. Following Refs. \cite{han02, Balestro03}, the quantum escape rate
\begin{equation}
\label{eq:gamma_MQT}
	\Gamma_{\mathrm{MQT}} =\frac{\omega_p}{2\pi} \exp\left[-\frac{36\Delta U}{5\hbar\omega_{p}} \left(1  +\frac{0.87\omega_p}{\omega_c}  \right)\right]
\end{equation}
involves again the $\theta$-dependent potential height $\Delta U$ and plasma frequency $\omega_p$ from Ref. \cite{lefevredevoret92}, with the angle $\theta$ close to $\pi/2$ now describing a trajectory dominated by the small mass $C_2$. The second term in the exponent of Eq.\,(\ref{eq:gamma_MQT}) takes into account the non-negligible effect of dissipation \cite{martinis87, CL83} (for the classical junction the ratio $\omega_p/\omega_c$ is small, allowing us to neglect damping \cite{pekola10}). Since instanton splitting \cite{1986_AnnPhy_Schmid, ivlev87} is not possible in the case of a capacitively asymmetric dc-SQUID \cite{Ivlev-PRB-2010} the additional prefactor $f_\mathrm{2D}$ used in Refs.\,\cite{han02, Balestro03} is set to unity and omitted in Eq.\,(\ref{eq:gamma_MQT}).

Equation\,(\ref{eq:gamma_MQT}) is used to fit the measured escape rates at different flux values (black rectangle in Fig.\,\ref{fig:flux_temp}) on the negative slope. The result of this fit and the corresponding measured escape rates at $\Phi_{\rm ext}/\Phi_0 = 0.24$ are shown in Fig.\,\ref{fig:fitratesTA}(b) for the three lowest temperatures. The deviation of the 270\,mK line is most probably due to the onset of thermally assisted tunneling. The resulting values for the fit parameters mean critical current $I_0$ and asymmetry parameter $\alpha$ for the two lowest temperatures agree within 4\% with the results obtained from the TA fit. The value for the resistance $R = 407\pm 26\,\Omega$ is decreased compared to the zero frequency value given in Tab.\,1. We attribute this discrepancy to an increased damping at high frequencies due to the smallness of our junction. In fact, we have observed the related effect of classical phase diffusion \cite{martiniskautz90} for a similar dc-SQUID ($L = 8\,$nH) with nominally the same junction sizes, fabricated on the same wafer. A large damping is also consistent with a reduced crossover temperature which lies between $ 270\, {\rm mK} < T_{\mathrm{cr}} < 430$\,mK according to Fig.\,4. This should be compared with the nominal crossover temperature at the switching current $T_{\rm cr} (j_{\rm sw})$ which is approximately 660\,mK. 

\begin{figure}[bt]
	\includegraphics[width =.9\columnwidth]{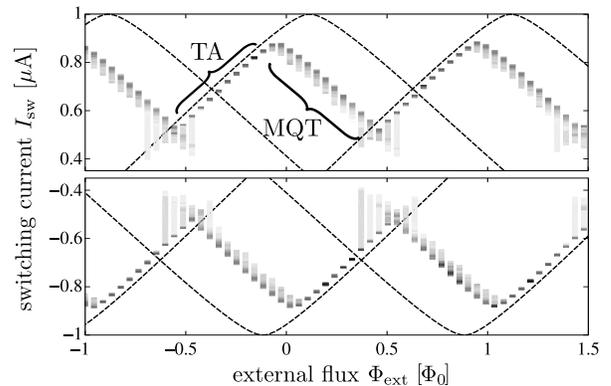}
\caption{Dependence of the switching current $I_{\rm sw}$ on flux $\Phi_{\rm ext}$ for positive and negative current polarity. The bars represent the recorded switching current histograms. The intensity of gray illustrates the number of counts per current bin. Solid lines display the theoretically expected behavior in the absence of fluctuations for the extracted critical currents.}
\label{fig:flux_base}
\end{figure}

Comparing TA and MQT rates for the two lowest temperatures shows that the escape statistics of the latter are much broader. This is further underlined in Fig.\,\ref{fig:flux_base} showing a measurement at lowest temperature over several periods of $\Phi_0$.
Instead of plotting only the mean switching current at the corresponding flux value, the full switching histogram is shown. The height of the histogram in every current bin with a finite number of events is presented in gray scale. Darker gray corresponds to a larger number of switching events in the respective bin.
For flux values in a small range around odd integer multiples of $\Phi_0/2$ the system has a non-negligible probability of being retrapped in a side minimum instead of the main one. Therefore, the measured histograms broaden and split into two peaks in this flux range. 
Figure \ref{fig:flux_base} shows clearly that in the asymmetric dc-SQUID the escape processes are different on the positive and negative slopes of $I_{\rm sw}(\Phi_{\rm ext})$. Measurements for positive and negative current polarity confirm that the observed MQT and TA behavior are indeed related to the escape direction.
Dashed black lines show again the calculated flux dependence of the critical currents for a classical dc-SQUID at zero temperature. The switching current on the MQT slope is considerably suppressed compared to the calculated critical current. The virtual particle escapes out of a minimum which, classically, it could not leave, yet.

In conclusion, we investigated the switching characteristics of a dc-SQUID with a strong capacitive asymmetry. Flux dependent measurements show a clear difference of the switching current distribution on positive and negative slopes. Temperature dependent measurements and the comparison with theory attribute the switching on positive slopes to TA and on negative slopes to MQT. Thus, depending on the magnetic field, our device shows either MQT or TA, both at the same temperature. Returning to the virtual particle picture, we showed that a particle with strongly anisotropic mass displays either quantum mechanical or classical behavior, depending on the direction of escape. Hence, coupling a classical to a quantum mechanical degree of freedom can facilitate the escape of a particle in a 2D potential considerably compared to a purely classical system.  We thus experimentally verified the predictions made by Thomann \itshape et al. \upshape \cite{Thomann09} and, on a less serious note, conclude that Baron M\"unchhausen could have pulled himself and his horse out of the swamp.

The authors acknowledge interesting discussions with B. I. Ivlev. Susanne Butz acknowledges the financial support by the Landesgraduiertenf\"orderung Baden-W\"urttemberg.

\bibliographystyle{unsrt}

\begin{thebibliography}{9}

\bibitem{Thomann09} A. U. Thomann, V. B. Geshkenbein and G. Blatter, Phys. Rev. B \textbf{79}, 184515 (2009).
 
\bibitem{CL83} A. O. Caldeira and A. Leggett, Ann. Phys. \textbf{149}, 374 (1983).

\bibitem{LA83} A. I. Larkin and Yu. N. Ovchinnikov, Zh. Eksp. Teor. Fiz. \textbf{85},
1510 (1983) [Sov. Phys. JETP \textbf{58}, 876 (1983)].

\bibitem{JJ65} B. D. Josephson, Advances in Physics \textbf{14}, 419 (1965).

\bibitem{DMC85} M. H. Devoret, J. M. Martinis, and J. Clarke, Phys. Rev. Lett. \textbf{55}, 1908 (1985).

\bibitem{Rudolph-APL-2012} M. Rudolph, J. Nagel, J. M. Meckbach, M. Kemmler, M. Siegel, K. Ilin, D. Koelle, and R. Kleiner, Appl. Phys. Lett. \textbf{101}, 052602 (2012).

\bibitem{Sullivan-JAP-2013} D. F. Sullivan, S. K. Dutta, M. Dreyer, M. A. Gubrud, A. Roychowdhury, J. R. Anderson, C. J. Lobb, and F. C. Wellstood, J. Appl. Phys. \textbf{113}, 183905 (2013).

\bibitem{Vion-Science-2002} D. Vion, A. Aassime, A. Cottet, P. Joyez, H. Pothier, C. Urbina, D. Esteve, and M. H. Devoret, Science \textbf{296}, 886 (2002).

\bibitem{Della-Rocca-PRL-2007} M. L. Della Rocca, M. Chauvin, B. Huard, H. Pothier, D. Esteve, and C. Urbina, Phys. Rev. Lett. \textbf{99}, 127005 (2007).

\bibitem{lefevredevoret92} V. Lefevre-Seguin, E. Turlot, C. Urbina, D. Esteve and M. H. Devoret, Phys. Rev. B \textbf{46}, 5507 (1992).

\bibitem{han02} S.-X. Li, Y. Yu, Y. Zhang, W. Qiu, S. Han and Z. Wang, Phys. Rev. Lett.\textbf{89}, 098301 (2002).

\bibitem{dolata05} R. Dolata and H. Scherer and A. B. Zorin and J. Niemeyer, J. Appl. Phys. \textbf{97}, 054501 (2005).

\bibitem{tinkham} M. Tinkham, {\it Introduction to superconductivity} (Dover Publ., Mineola, NY, 2004), p.261.


\bibitem{wallraff03} A. Wallraff, A. Lukashenko, C. Coqui, A. Kemp, T. Duty and A. V. Ustinov, Rev. Scient. Inst. \textbf{74}, 3740 (2003).

\bibitem{fulton74} T. A. Fulton and L. N. Dunkleberg, Phys. Rev. B \textbf{9}, 4760 (1974)

\bibitem{Balestro03} F. Balestro, J. Claudon, J. P. Pekola and O. Buisson, Phys. Rev. Lett. \textbf{91}, 158301, (2003).

\bibitem{martinis87} J. M. Martinis, M. H. Devoret and J. Clarke, Phys. Rev. B \textbf{35}, 4682 (1987).

\bibitem{pekola10} Y. Yoon, S. Gasparinetti, M. M\"ott\"onen and J. P. Pekola, J. Low Temp. Phys. \textbf{163}, 164 (2011).

\bibitem{1986_AnnPhy_Schmid} A. Schmid, Ann. Phys. \textbf{170}, 33 (1986).

\bibitem{ivlev87} B. I. Ivlev and Y. N. Ovchinnikov, Zh. Eksp. Teor. Fiz. \textbf{93}, 668-679 (1987).

\bibitem{Ivlev-PRB-2010} B. Ivlev and J. P. Palomares-Baez, Phys. Rev. B \textbf{82}, 184513 (2010).


\bibitem{martiniskautz90} R. L. Kautz and J. M. Martinis, Phys. Rev. B \textbf{42}, 9903 (1990).







%\bibitem{Lecocq-PRL-2012} F. Lecocq, I. M. Pop, I. Matei, E. Dumur, A.K. Feofanov, C. Naud, W. Guichard, and O. Buisson, Phys. Rev. Lett.  \textbf{108}, 107001 (2012).

%\bibitem{martinisclarke87} J. M. Martinis, M. H. Devoret and J. Clarke, Phys. Rev. B \textbf{35}, 10 (1987).






\end{thebibliography}

\end{document}